\documentclass[conference]{IEEEtran}
\IEEEoverridecommandlockouts

\usepackage{cite}
\usepackage{amsmath,amssymb,amsfonts}
\usepackage{algorithmic}
\usepackage{graphicx}
\usepackage{textcomp}
\usepackage{xcolor}
\usepackage{graphicx}
\usepackage[caption=false,font=footnotesize]{subfig}
\def\BibTeX{{\rm B\kern-.05em{\sc i\kern-.025em b}\kern-.08em
    T\kern-.1667em\lower.7ex\hbox{E}\kern-.125emX}}
\begin{document}

\title{Measurement-Based FR3 Urban Macrocell Channel Characterization and Coverage Analysis\\

}

\author{
\IEEEauthorblockN{
Enrui Liu\IEEEauthorrefmark{1},
Pan Tang\IEEEauthorrefmark{1},
Haiyang Miao\IEEEauthorrefmark{2},
Qi Zhen\IEEEauthorrefmark{1},
and Jianhua Zhang\IEEEauthorrefmark{1}
}
\IEEEauthorblockA{
\IEEEauthorrefmark{1}Beijing University of Posts and Telecommunications, Beijing, China\\
Email: \{liuenrui, tangpan27, zq2024018002, jhzhang\}@bupt.edu.cn\\
\IEEEauthorrefmark{2}Tsinghua University, Beijing, China\\
Email: miaohy@tsinghua.edu.cn
}
}

\maketitle

\begin{abstract}
The 6--18~GHz upper mid-band is a promising spectrum range for future sixth-generation (6G) networks due to its favorable balance between available bandwidth and propagation capability, while its same-site coverage performance remains a key deployment concern. This paper presents a wideband channel measurement campaign at 13 spatially aligned frequency points from 6 to 18~GHz in an urban macrocell (UMa) environment, and investigates the frequency evolution of channel characteristics and same-site coverage capability. The results show that path loss generally increases with frequency, while the higher-frequency channels tend to exhibit smaller root-mean-square (RMS) delay spread (DS) and larger PDP-based $K$-factors. Based on the measured path-loss characteristics, the additional link-budget requirement and service-aware same-site coverage are further evaluated. Under the baseline service configuration, the 80\% coverage distance $d_{80}$ decreases from approximately 129~m at 6~GHz to 58~m at 18~GHz. Further analysis quantifies the impacts of service rate, resource allocation, interference, and link-budget degradation on coverage, providing measurement-based insights for same-site deployment across the 6--18~GHz band.
\end{abstract}


\section{Introduction}

With the growing demand for wide bandwidth and high-capacity wireless access in sixth-generation (6G) networks, the upper mid-band spectrum from 6 to 18~GHz has emerged as a promising candidate due to its abundant spectrum resources and more favorable propagation characteristics than millimeter-wave bands\cite{1}. However, increasing the carrier frequency generally leads to higher propagation loss and stronger sensitivity to the surrounding environment\cite{FR3}, potentially diminishing the coverage advantage offered by conventional lower-frequency networks\cite{2}. From a practical deployment perspective, whether existing macro-cell sites can be reused without substantially increasing site density is therefore a critical issue for the large-scale deployment of upper mid-band networks\cite{Bjornson1}. Consequently, beyond characterizing the frequency-dependent propagation behavior, it is essential to quantify the additional link budget required when migrating from lower to higher frequencies and to determine how the coverage range evolves under the same-site deployment and service constraints.

Existing studies on wireless propagation above 6~GHz can be broadly divided into two categories. The first category investigates the frequency dependence of channel characteristics, such as path loss (PL)\cite{haiyang2}, shadow fading, root-mean-square (RMS) delay spread (DS)\cite{haiyang3}, and Ricean $K$-factor, through multi-frequency channel measurements\cite{haiyang, JCIN}.However, most studies consider only a limited number of discrete frequency bands\cite{Rappaport}, while measurement locations and propagation environments are not always strictly aligned across frequencies, making it difficult to isolate the frequency-dependent evolution of the channel. The second category evaluates coverage performance across different frequency bands based on link-budget analysis, system-level simulations, or standardized channel models\cite{liu, jinshi2026}. Existing coverage evaluations are largely based on standardized models, while comprehensive same-site measurements across the mid-band remain limited, leaving the translation from measured frequency-dependent propagation variations to cross-frequency link-budget requirements and same-site coverage insufficiently explored.

To address these issues, this paper conducts a multi-frequency wideband channel measurement campaign from 6 to 18~GHz in an urban macrocell (UMa) scenario. The main contributions are summarized as follows:

\begin{itemize}

\item A wideband channel measurement campaign is conducted at 13 frequency points from 6 to 18~GHz in a UMa scenario with spatially aligned measurement locations.

\item The frequency evolution of PL, RMS DS, and PDP-based $K$-factor is characterized, revealing the propagation behavior across the 6--18~GHz band.

\item Measurement-based same-site coverage is evaluated in terms of cross-frequency link-budget requirements, coverage evolution, and the impacts of service and system constraints.

\end{itemize}

\section{Channel Measurement Platform and Environment}
\subsection{Measurement Platform}

\begin{figure}[!hbt]
    \centering
    \includegraphics[width=\columnwidth]{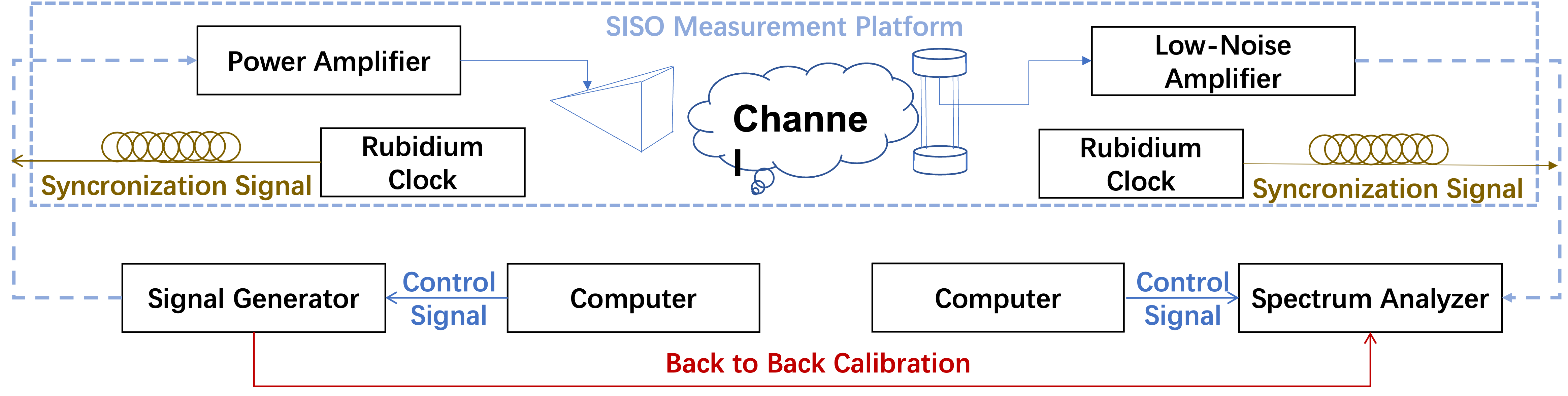}
    \caption{Architecture of the SISO channel measurement platform.}
    \label{fig:siso_platform}
\end{figure}

The SISO channel measurement platform is illustrated in Fig.~\ref{fig:siso_platform}, and the main measurement parameters are summarized in Table~\ref{tab:measurement_parameters}. The platform consists of a transmitter (Tx), a receiver (Rx), wideband antennas, synchronization modules, and control units, supporting wideband channel sounding at multiple carrier frequencies from 6 to 18~GHz. At the Tx, a PN9 sounding signal is generated and transmitted through a wideband horn antenna. After propagation through the wireless channel, the signal is captured by a biconical antenna at the Rx and recorded for subsequent channel impulse response (CIR) estimation\cite{steinbauer2001double}. The Tx and Rx are coordinated through synchronization and control signals to ensure consistent data acquisition during the measurements. The detailed sounding configuration, antenna parameters, and measurement settings are provided in Table~\ref{tab:measurement_parameters}.

\begin{table}[!hbt]
\centering
\caption{SISO Channel Measurement Parameters}
\label{tab:measurement_parameters}
\renewcommand{\arraystretch}{1.1}
\begin{tabular}{l c}
\hline
\textbf{Parameter} & \textbf{Value} \\
\hline
Measurement scenario              & UMa \\
Carrier frequencies               & 6--18 GHz \\
Frequency interval                & 1 GHz \\
Signal bandwidth                  & 400 MHz \\
Sounding sequence                 & PN9 \\
Sampling rate                     & 400 MSa/s \\
Transmit power                    & 10 dBm \\
Tx antenna type                   & Wideband horn antenna \\
Rx antenna type                   & Biconical antenna \\
Typical Tx antenna gain           & 12 dBi \\
Typical Rx antenna gain           & 3 dBi \\
Tx antenna height                 & 27.8 m \\
Rx antenna height                 & 1.6 m \\
Tx/Rx polarization                & V--V \\
Number of measurement routes      & 4 \\
Rx point spacing                  & $\sim$20 m \\
Snapshots per measurement point  & 100 \\
Calibration method                & Back-to-back calibration \\
\hline
\end{tabular}
\end{table}

\subsection{Measurement Environment}
The measurement campaign was conducted on the Shahe Campus of Beijing University of Posts and Telecommunications (BUPT). For the UMa scenario, the transmitter (Tx) was deployed on the rooftop of the BUPT main building at a height of approximately 27.8~m above ground, as marked by the red star in Fig.~\ref{fig:siso_platform}(a). The receiver (Rx) was moved along the four measurement routes shown in Fig.~\ref{fig:siso_platform}(a), with adjacent measurement points spaced approximately 20~m apart. Both line-of-sight (LOS) and non-line-of-sight (NLOS) propagation conditions were considered, as indicated by the white and yellow circles, respectively. At each measurement point, the Rx remained stationary during data acquisition to ensure quasi-static channel conditions. Fig.~\ref{fig:siso_platform}(b) presents a photograph of the UMa measurement environment, showing the locations of the rooftop Tx and the ground-level Rx.
\begin{figure}[!t]
    \centering
    \includegraphics[width=0.48\columnwidth]{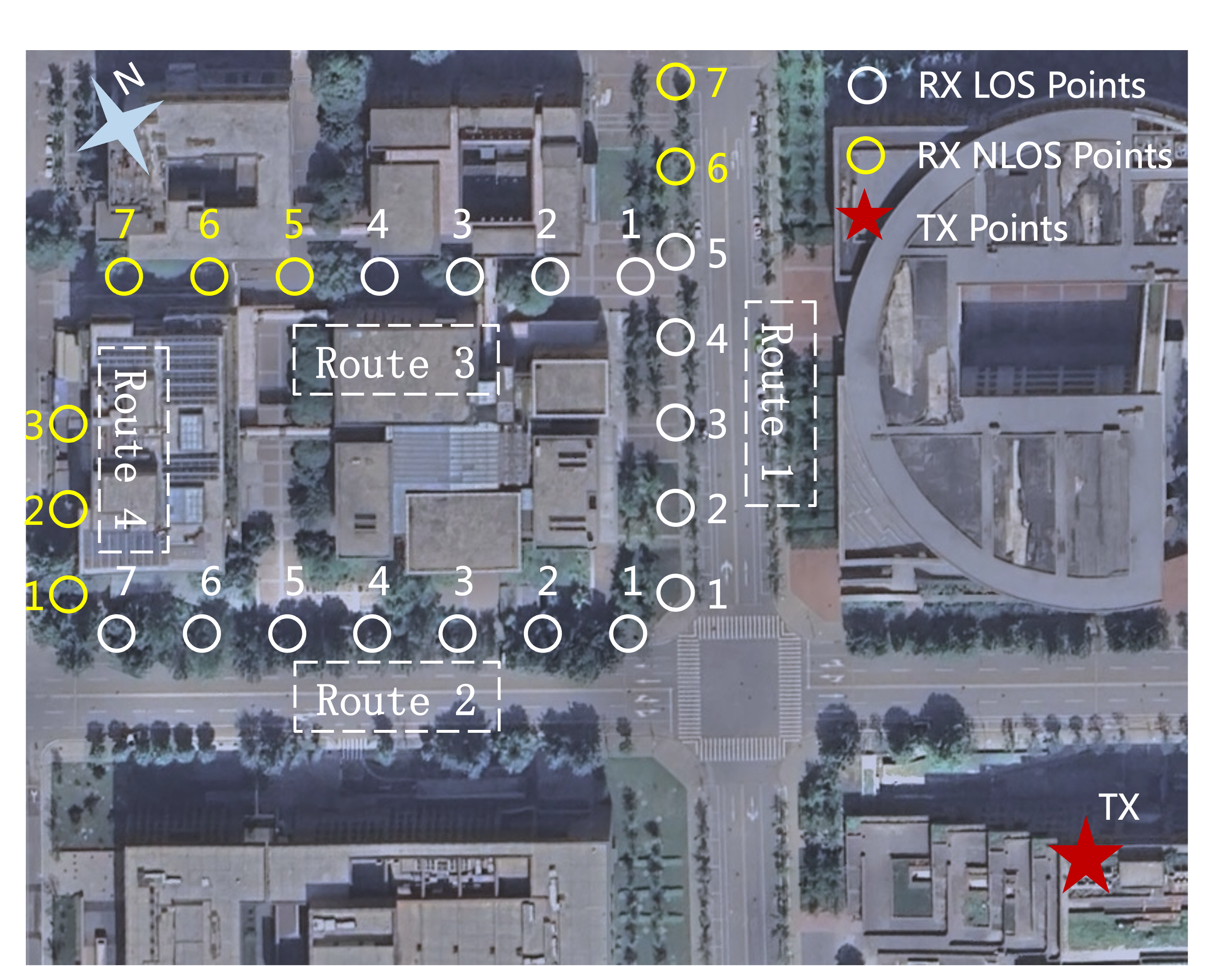}
    \hfill
    \includegraphics[width=0.48\columnwidth]{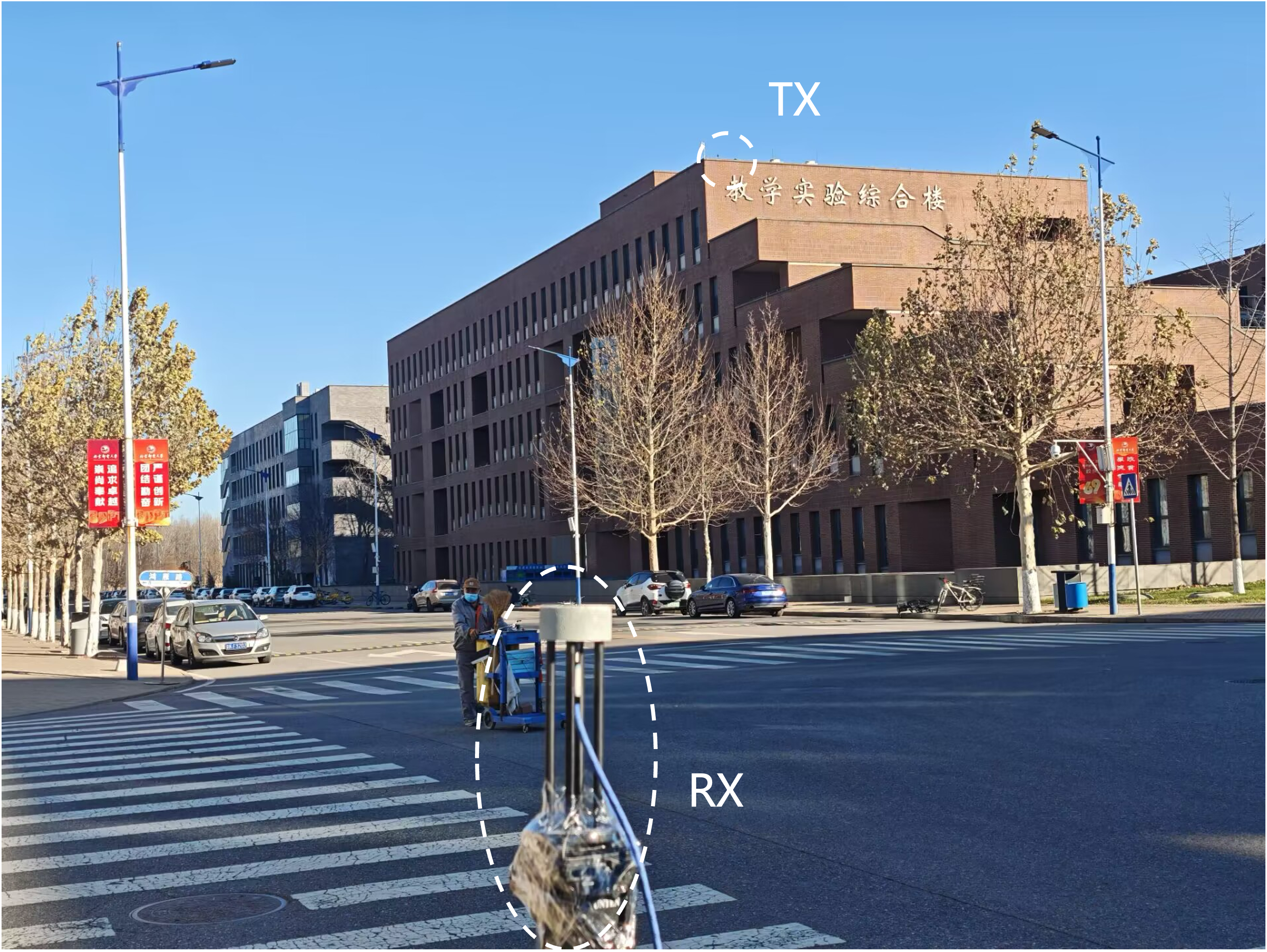}
    \caption{UMa measurement scenario: 
    (a) measurement routes and Rx locations; 
    (b) measurement environment.}
    \label{fig:measurement_scenario}
\end{figure}

\subsection{Measurement Data Post-Processing}
A PN9 sequence is employed as the sounding waveform due to its favorable autocorrelation property, which facilitates reliable extraction of the CIR. To remove the inherent response of the measurement system, a back-to-back calibration is performed prior to the measurement campaign \cite{wang,tangpan}. The received signal can be expressed as
\begin{equation}
y(\tau)
=
x(\tau)
*h_{\mathrm{Tx}}(\tau)
*a_{\mathrm{Tx}}(\tau)
*h(\tau)
*a_{\mathrm{Rx}}(\tau)
*h_{\mathrm{Rx}}(\tau),
\end{equation}
where $*$ denotes convolution, $x(\tau)$ is the transmitted PN sequence, $h_{\mathrm{Tx}}(\tau)$ and $h_{\mathrm{Rx}}(\tau)$ represent the responses of the Tx and Rx RF chains, respectively, $a_{\mathrm{Tx}}(\tau)$ and $a_{\mathrm{Rx}}(\tau)$ denote the responses of the Tx and Rx antennas, and $h(\tau)$ is the propagation-channel impulse response.

During back-to-back calibration, the Tx and Rx RF chains are directly connected through a cable while bypassing the wireless channel and antennas. The corresponding calibration signal is given by
\begin{equation}
y_{\mathrm{cal}}(\tau)
=
x(\tau)
*h_{\mathrm{Tx}}(\tau)
*h_{\mathrm{Rx}}(\tau).
\end{equation}
Accordingly, the frequency-dependent responses of the Tx and Rx RF chains can be removed by frequency-domain deconvolution. The calibrated response is obtained as
\begin{equation}
h_{\mathrm{cal}}(\tau)
=
\mathcal{F}^{-1}
\left\{
\frac{Y(f)}{Y_{\mathrm{cal}}(f)}
\right\}
=
a_{\mathrm{Tx}}(\tau)
*h(\tau)
*a_{\mathrm{Rx}}(\tau),
\end{equation}
where $\mathcal{F}^{-1}\{\cdot\}$ denotes the inverse Fourier transform, and $Y(f)$ and $Y_{\mathrm{cal}}(f)$ are the frequency-domain representations of the received and calibration signals, respectively.

This calibration compensates for both the RF-chain and antenna responses, which is particularly important for consistent cross-frequency comparison over the 6--18~GHz band. The resulting calibrated CIRs are subsequently used to extract the path loss, RMS delay spread, and PDP-based $K$-factor.

\section{Channel Characteristics}
\subsection{Path Loss}

Path loss (PL) is first investigated to characterize the large-scale propagation behavior over the 6--18~GHz band. Based on the calibrated CIR, the received power at the $i$-th measurement location is calculated by integrating the power of the effective multipath components as
\begin{equation}
P_{\mathrm{r}}(f,d_i)
=
\sum_{\tau \in \Omega_i}
\left|h(f,d_i,\tau)\right|^2,
\end{equation}
where $h(f,d_i,\tau)$ denotes the calibrated CIR at carrier frequency $f$ and T-R separation distance $d_i$, and $\Omega_i$ denotes the set of valid delay bins after noise removal. Accordingly, the measured PL is obtained as
\begin{equation}
PL(f,d_i)
=
P_{\mathrm{t}}
-10\log_{10}\left(P_{\mathrm{r}}(f,d_i)\right),
\end{equation}
where $P_{\mathrm{t}}$ is the transmit power. To characterize the distance dependence of PL at each frequency, the close-in (CI) reference-distance model is adopted as
\begin{equation}
PL^{\mathrm{CI}}(f,d)
=
FSPL(f,d_0)
+
10n(f)\log_{10}\left(\frac{d}{d_0}\right)
+
X_{\sigma}(f),
\label{eq:CI}
\end{equation}
where $d_0=1$~m, $n(f)$ is the path loss exponent (PLE), and $X_{\sigma}(f)$ is a zero-mean Gaussian random variable with a standard deviation of $\sigma(f)$, representing shadow fading. The free-space path loss at the reference distance is given by
\begin{equation}
FSPL(f,d_0)
=
20\log_{10}
\left(
\frac{4\pi f d_0}{c}
\right),
\end{equation}
where $c$ denotes the speed of light. The PLE is estimated independently for LOS and NLOS conditions at each carrier frequency by minimizing the squared fitting error between the measured PL and the CI model.

\begin{figure}[!t]
    \centering
    \includegraphics[width=\columnwidth]{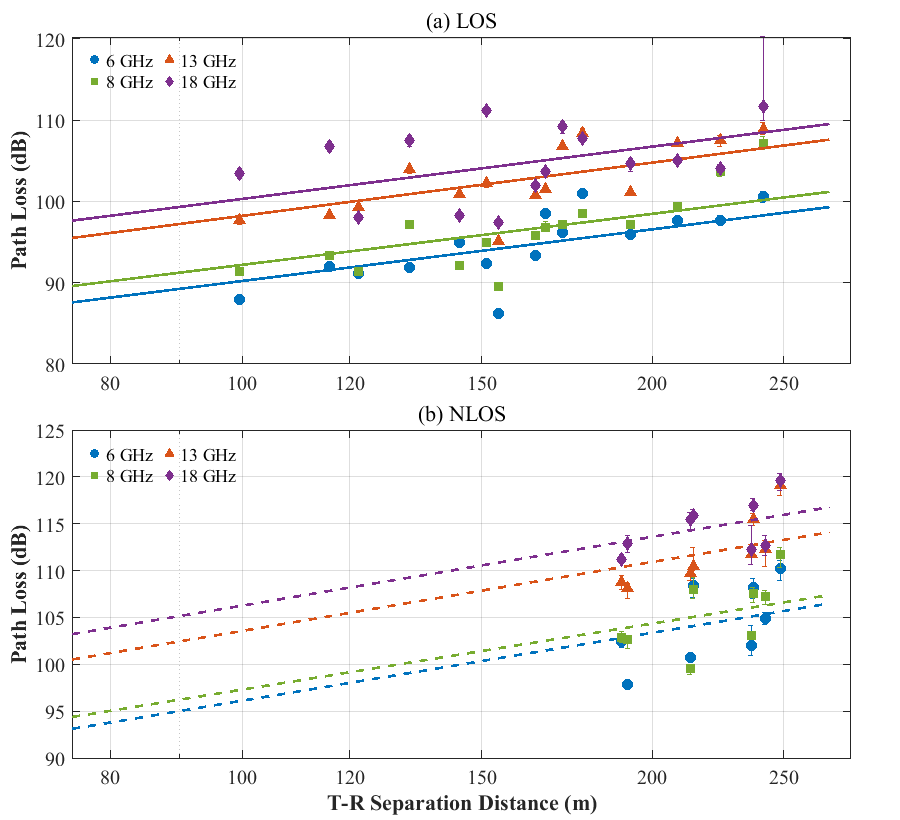}
    \caption{Measured path loss and CI model fitting at representative 
    frequencies under (a) LOS and (b) NLOS conditions.}
    \label{fig:PL}
\end{figure}

\begin{figure}[!t]
    \centering
    \includegraphics[width=0.9\columnwidth]{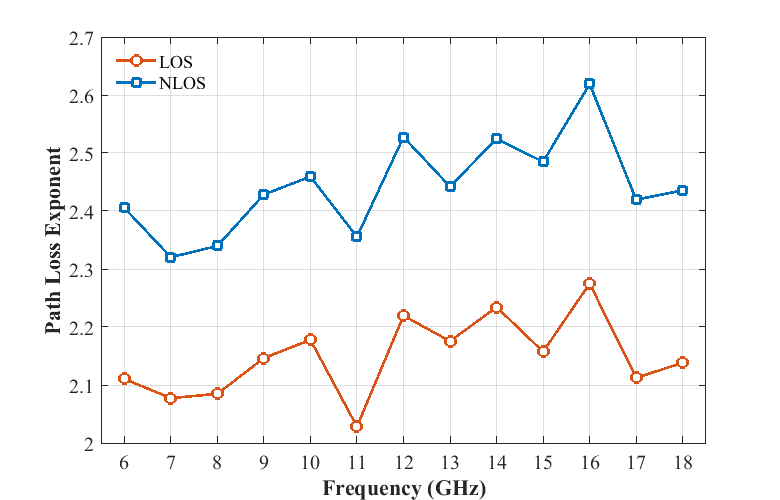}
    \caption{Path loss exponent versus carrier frequency under LOS and 
    NLOS conditions.}
    \label{fig:PLE}
\end{figure}

Fig.~\ref{fig:PL} shows the measured PL versus T-R separation distance at representative frequencies of 6, 8, 13, and 18~GHz. As expected, the PL generally increases with both T-R separation distance and carrier frequency. For LOS conditions, the measured samples are relatively concentrated around the fitted curves, indicating a stable distance-dependent attenuation trend. In contrast, the NLOS measurements exhibit larger fluctuations due to blockage and the stronger dependence on reflection and diffraction paths. A clear frequency-dependent offset can also be observed: higher-frequency links generally experience larger PL at comparable distances, while the overall distance-dependent slopes remain relatively similar within each propagation condition. To further examine the frequency dependence of the distance attenuation, Fig.~\ref{fig:PLE} presents the estimated PLEs from 6 to 18~GHz. The LOS PLE varies approximately from 2.03 to 2.28, while the NLOS PLE ranges from approximately 2.32 to 2.62. The NLOS PLE is consistently larger than its LOS counterpart, demonstrating the additional distance-dependent attenuation caused by obstruction and indirect propagation. Although the PLE does not increase monotonically with frequency, an overall upward tendency can be observed for both LOS and NLOS conditions. This indicates that the frequency-dependent PL increase is dominated by the FSPL term, accompanied by a moderate increase in distance-dependent attenuation.

\subsection{RMS Delay Spread}

The RMS-DS is employed to characterize the temporal dispersion of the measured channels. Based on the calibrated power delay profile (PDP), the RMS DS is calculated as
\begin{equation}
\tau_{\mathrm{RMS}}
=
\sqrt{
\frac{\sum_{l} P_l \tau_l^2}{\sum_{l} P_l}
-
\left(
\frac{\sum_{l} P_l \tau_l}{\sum_{l} P_l}
\right)^2
},
\label{eq:rms_ds}
\end{equation}
where $P_l$ and $\tau_l$ denote the power and excess delay of the $l$-th effective multipath component, respectively. Following the conventional statistical characterization of delay spread, the RMS DS is analyzed in the logarithmic domain as
\begin{equation}
\lg DS = \log_{10}\left(\frac{\tau_{\mathrm{RMS}}}{1~\mathrm{s}}\right),
\label{eq:lgds}
\end{equation}
and is modeled by a Gaussian distribution,
\begin{equation}
\lg DS \sim \mathcal{N}\left(\mu_{\mathrm{DS}},
\sigma_{\mathrm{DS}}^2\right),
\label{eq:ds_gaussian}
\end{equation}
where $\mu_{\mathrm{DS}}$ and $\sigma_{\mathrm{DS}}$ denote the mean and standard deviation of $\lg DS$, respectively.

\begin{figure}[!t]
    \centering
    \includegraphics[width=\columnwidth]
    {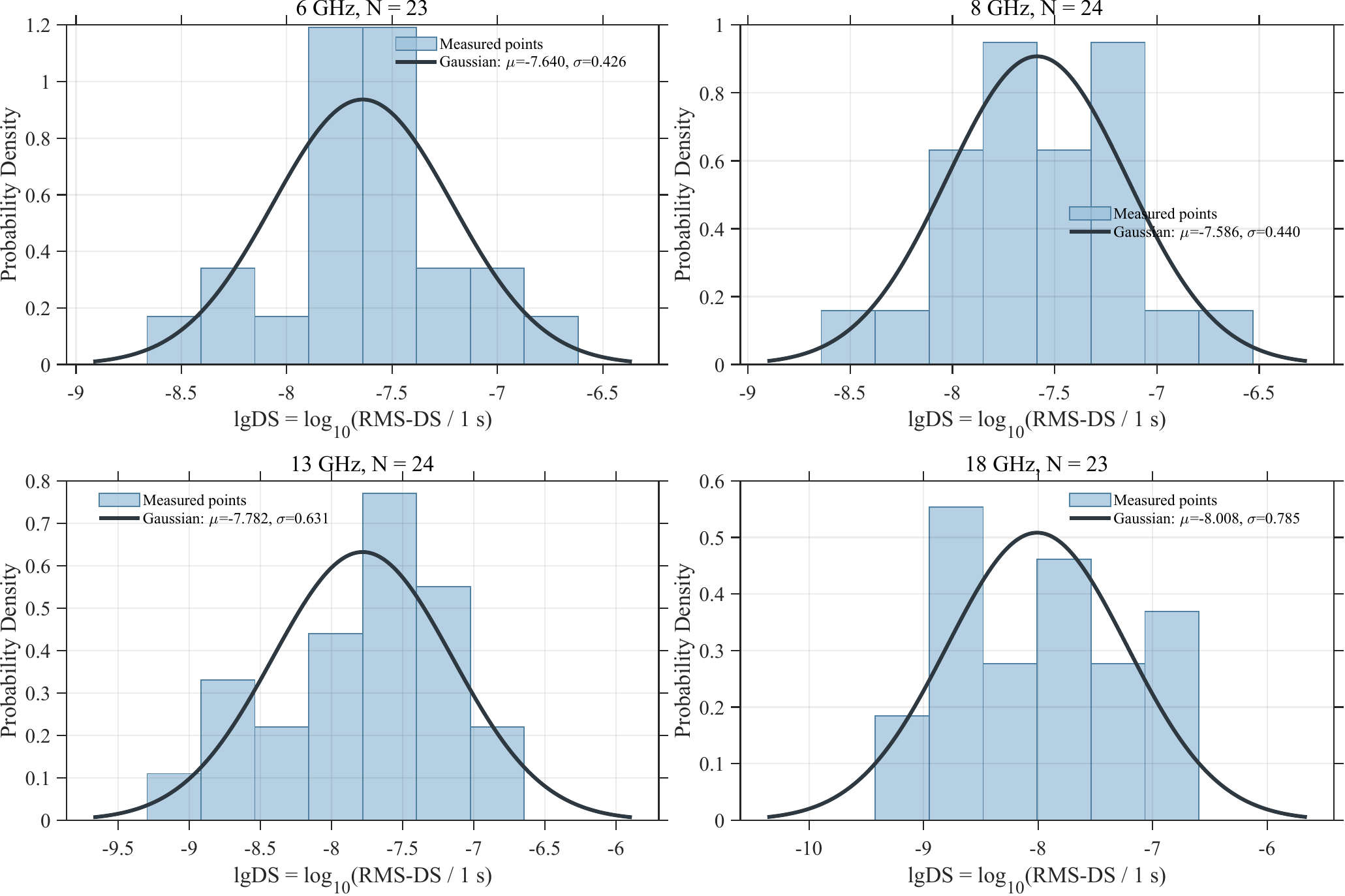}
    \caption{Probability distributions and Gaussian fittings of logarithmic 
    RMS DS at representative frequencies.}
    \label{fig:ds_hist}
\end{figure}

Fig.~\ref{fig:ds_hist} presents the distributions of $\lg DS$ and their Gaussian fittings at representative frequencies of 6, 8, 13, and 18~GHz. The fitted means are $-7.640$, $-7.586$, $-7.782$, and $-8.008$, respectively, while the corresponding standard deviations are $0.426$, $0.440$, $0.631$, and $0.785$. The results show that the mean delay dispersion generally decreases toward the higher-frequency bands, particularly at 18~GHz. This behavior can be attributed to the larger attenuation of reflected and diffracted components at higher frequencies, which reduces the contribution of weak multipath components with long excess delays. Meanwhile, the larger $\sigma_{\mathrm{DS}}$ observed at 13 and 18~GHz indicates greater variability of delay dispersion. These results suggest that increasing the carrier frequency does not simply shift the delay-spread distribution, but also changes its spatial variability across the measured UMa environment.

\subsection{PDP-based $K$-Factor}

The PDP-based $K$-factor is further investigated to characterize the relative strength of the dominant propagation component with respect to the remaining multipath components. Based on the measured PDP, the $K$-factor is calculated as
\begin{equation}
K
=
10\log_{10}
\left(
\frac{P_{\mathrm{dom}}}
{P_{\mathrm{tot}}-P_{\mathrm{dom}}}
\right),
\label{eq:kfactor}
\end{equation}
where $P_{\mathrm{dom}}$ denotes the power of the dominant multipath component and $P_{\mathrm{tot}}$ is the total received multipath power. A larger $K$-factor indicates that the received signal is more strongly dominated by a single propagation component, whereas a smaller value corresponds to a richer multipath environment.

\begin{figure}[!t]
    \centering
    \includegraphics[width=0.9\columnwidth]{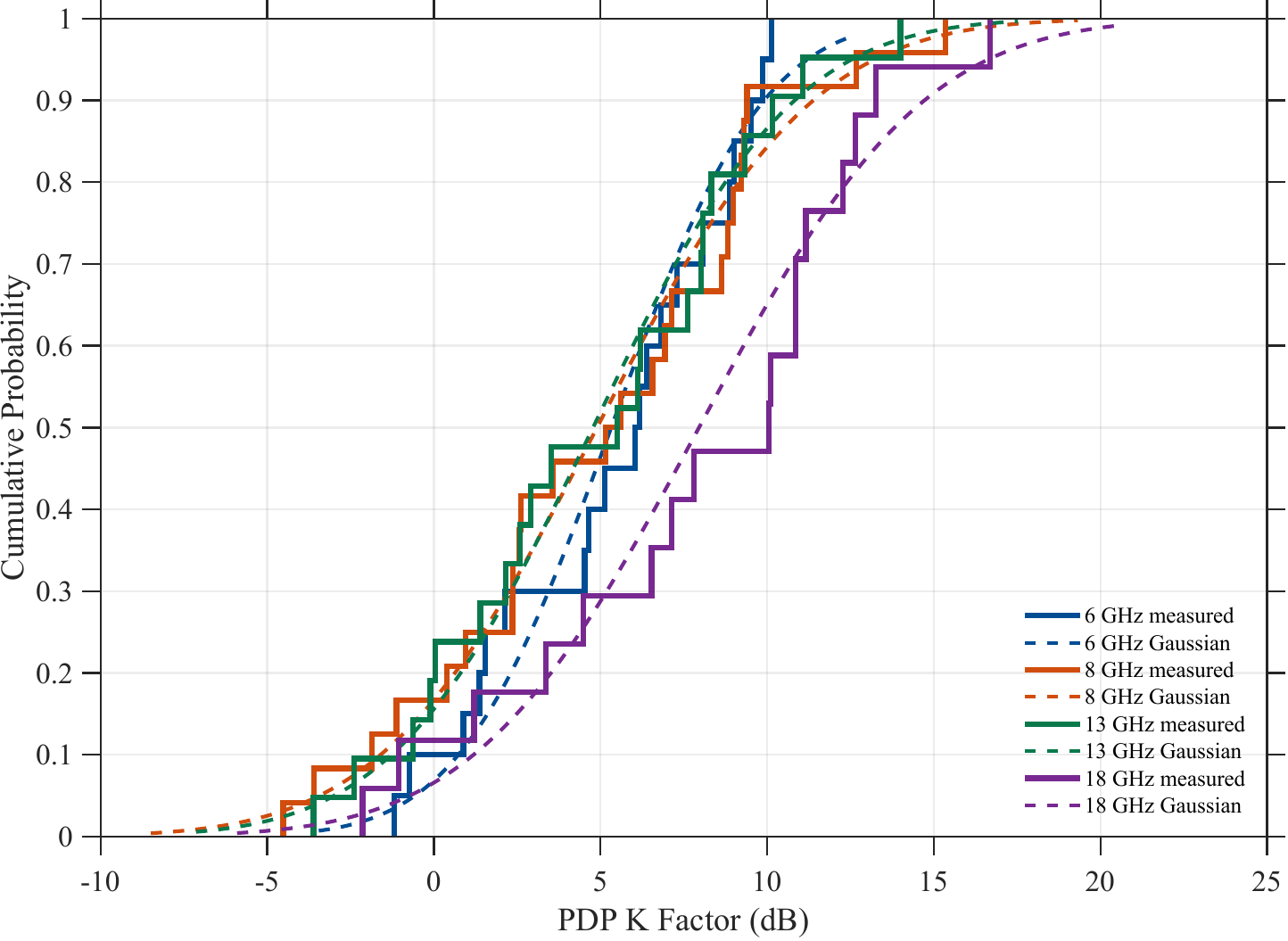}
    \caption{Empirical and Gaussian-fitted CDFs of the PDP-based $K$-factor 
    at representative frequencies.}
    \label{fig:k_ecdf}
\end{figure}

The cumulative distribution functions (CDF) of the $K$-factor at representative frequencies are shown in Fig.~\ref{fig:k_ecdf}, together with their Gaussian fittings. The distributions at 6, 8, and 13~GHz are relatively close, indicating comparable dominant-to-diffuse power ratios over these frequencies. In contrast, the distribution at 18~GHz is clearly shifted toward larger $K$-factor values, suggesting a stronger dominance of the main propagation component at the higher frequency.

This observation is also consistent with the RMS DS results. As the carrier frequency increases, weak reflected and diffracted components generally experience stronger attenuation, reducing their contributions to the received multipath power. Consequently, the received power tends to become more concentrated in the dominant component, resulting in a larger $K$-factor and, simultaneously, a smaller delay spread. The joint behavior of RMS DS and $K$-factor therefore indicates a tendency toward more dominant-component-driven propagation at the upper end of the measured 6--18~GHz band.

\section{Same-Site Coverage Analysis}

Based on the measured channel characteristics over 6--18~GHz, this section investigates cross-frequency link-budget requirements and same-site coverage capability under different service and system constraints.

\subsection{Service-Aware Same-Site Coverage Criterion}

To translate the measured propagation characteristics into service coverage, a service-aware link-budget criterion is first established. For a target net user rate $R_{\mathrm{net}}$, the required spectral efficiency is given by
\begin{equation}
\eta_{\mathrm{req}}
=
\frac{R_{\mathrm{net}}}
{B_{\mathrm{sys}}\rho_{\mathrm{sch}}(1-\eta_{\mathrm{OH}})},
\label{eq:required_se}
\end{equation}
where $B_{\mathrm{sys}}$ is the system bandwidth, $\rho_{\mathrm{sch}}$ is the fraction of radio resources allocated to the user, and $\eta_{\mathrm{OH}}$ denotes the protocol and signaling overhead.

Based on the Shannon bound\cite{capacity}, the required SNR is
\begin{equation}
\gamma_{\mathrm{Sh}}
=
2^{\eta_{\mathrm{req}}}-1.
\label{eq:shannon_snr}
\end{equation}
Considering the implementation and coding gap $G_{\mathrm{imp}}$ and the additional margin $M_{\mathrm{BLER}}$ for the target block error rate, the required SINR in dB is expressed as
\begin{equation}
\gamma_{\mathrm{req,dB}}
=
10\log_{10}\left(\gamma_{\mathrm{Sh}}\right)
+
G_{\mathrm{imp}}
+
M_{\mathrm{BLER}}.
\label{eq:required_sinr}
\end{equation}
The receiver noise power is calculated as
\begin{equation}
N_{\mathrm{Rx}}
=
-174
+
10\log_{10}(B_{\mathrm{sys}})
+
NF,
\label{eq:receiver_noise}
\end{equation}
where $NF$ denotes the receiver noise figure. When interference is considered, the corresponding noise rise is
\begin{equation}
\Delta N_I
=
10\log_{10}
\left(
1+10^{(I/N)/10}
\right),
\label{eq:noise_rise}
\end{equation}
where $I/N$ denotes the interference-to-noise ratio. The minimum required received power and the corresponding available path-loss budget are therefore given by
\begin{equation}
P_{\mathrm{Rx,min}}
=
N_{\mathrm{Rx}}
+
\Delta N_I
+
\gamma_{\mathrm{req,dB}},
\label{eq:receiver_threshold}
\end{equation}
and
\begin{equation}
PL_{\mathrm{budget}}
=
P_{\mathrm{Tx}}
-
P_{\mathrm{Rx,min}},
\label{eq:pl_budget}
\end{equation}
respectively, where $P_{\mathrm{Tx}}$ is the transmit-power reference. To incorporate the spatial variation observed in the measurements, the frequency-dependent path loss is modeled as\cite{3gpp38901}
\begin{equation}
PL(f,d)
=
\alpha(f)
+
\beta(f)\log_{10}(d)
+
X_{\sigma}(f),
\label{eq:coverage_pl_model}
\end{equation}
where $\alpha(f)$ and $\beta(f)$ are the fitted parameters and $X_{\sigma}(f)\sim\mathcal{N}(0,\sigma^2(f))$ represents the spatial variation around the fitted model. The coverage model is fitted using all measurement points regardless of propagation condition. For a given path-loss budget, the spatial coverage probability at distance $d$ is then expressed as
\begin{equation}
P_{\mathrm{cov}}(f,d)
=
\Phi
\left[
\frac{
PL_{\mathrm{budget}}
-\alpha(f)
-\beta(f)\log_{10}(d)
}{
\sigma(f)
}
\right],
\label{eq:coverage_probability}
\end{equation}
where $\Phi(\cdot)$ denotes the standard normal cumulative distribution function. The detailed parameter settings are summarized in Table. \ref{tab:coverage_parameters}

\begin{table}[!t]
\centering
\caption{Baseline Parameters for Same-Site Coverage Evaluation}
\label{tab:coverage_parameters}
\renewcommand{\arraystretch}{1.08}
\setlength{\tabcolsep}{5pt}
\begin{tabular}{l c}
\hline
\textbf{Parameter} & \textbf{Value} \\
\hline
Reference carrier frequency, $f_{\mathrm{ref}}$     & 6 GHz \\
System bandwidth, $B_{\mathrm{sys}}$                & 400 MHz \\
Target net user rate, $R_{\mathrm{net}}$            & 50 Mb/s \\
Scheduled resource fraction, $\rho_{\mathrm{sch}}$  & 100\% \\
Protocol/signaling overhead, $\eta_{\mathrm{OH}}$   & 14\% \\
Receiver noise figure, $NF$                         & 7 dB \\
Implementation/coding gap, $G_{\mathrm{imp}}$       & 2 dB \\
BLER margin, $M_{\mathrm{BLER}}$                    & 1 dB \\
Interference-to-noise ratio, $I/N$                  & $-\infty$ dB \\
External loss                                       & 0 dB \\
Fast-fading margin                                  & 0 dB \\
Planning margin                                     & 0 dB \\
Transmit-power reference, $P_{\mathrm{Tx}}$         & 10 dBm \\
Spatial coverage probability                        & 80\% \\
\hline
\end{tabular}
\end{table}

\subsection{Cross-Frequency Coverage Evolution}

\begin{figure}[!t]
\centering
\includegraphics[width=\columnwidth]
{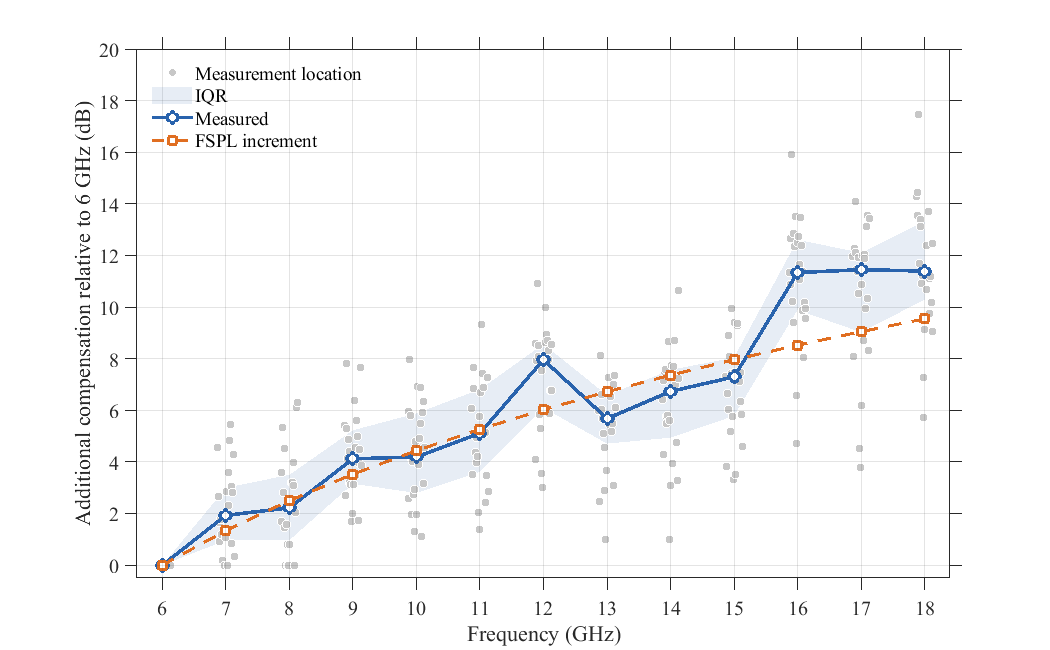}
\caption{Additional link-budget compensation relative to 6~GHz across the 
6--18~GHz band.}
\label{fig:compensation}
\end{figure}

We first quantify the additional link-budget requirement when migrating from 6~GHz toward higher frequencies. Owing to the spatial alignment of the multi-frequency measurements, the path-loss difference at the same location can be directly calculated as
\begin{equation}
\Delta PL_i(f)
=
PL_i(f)-PL_i(f_{\mathrm{ref}}),
\qquad
f_{\mathrm{ref}}=6~\mathrm{GHz},
\label{eq:delta_pl}
\end{equation}
where $PL_i(f)$ denotes the measured path loss at the $i$-th location and frequency $f$. The required additional link-budget compensation is defined as
\begin{equation}
C_i(f)
=
\max\left[\Delta PL_i(f),0\right].
\label{eq:compensation}
\end{equation}
Fig.~\ref{fig:compensation} shows the additional link-budget compensation relative to 6~GHz. The shaded region denotes the interquartile range (IQR) across measurement locations. The measured compensation generally increases with frequency and follows the FSPL increment, while considerable location-dependent variations are observed. In particular, the measured compensation at 16--18~GHz exceeds the corresponding FSPL increment, indicating additional environment-dependent propagation penalties beyond free-space frequency scaling.

\begin{figure}[!hbt]
\centering
\includegraphics[width=\columnwidth]
{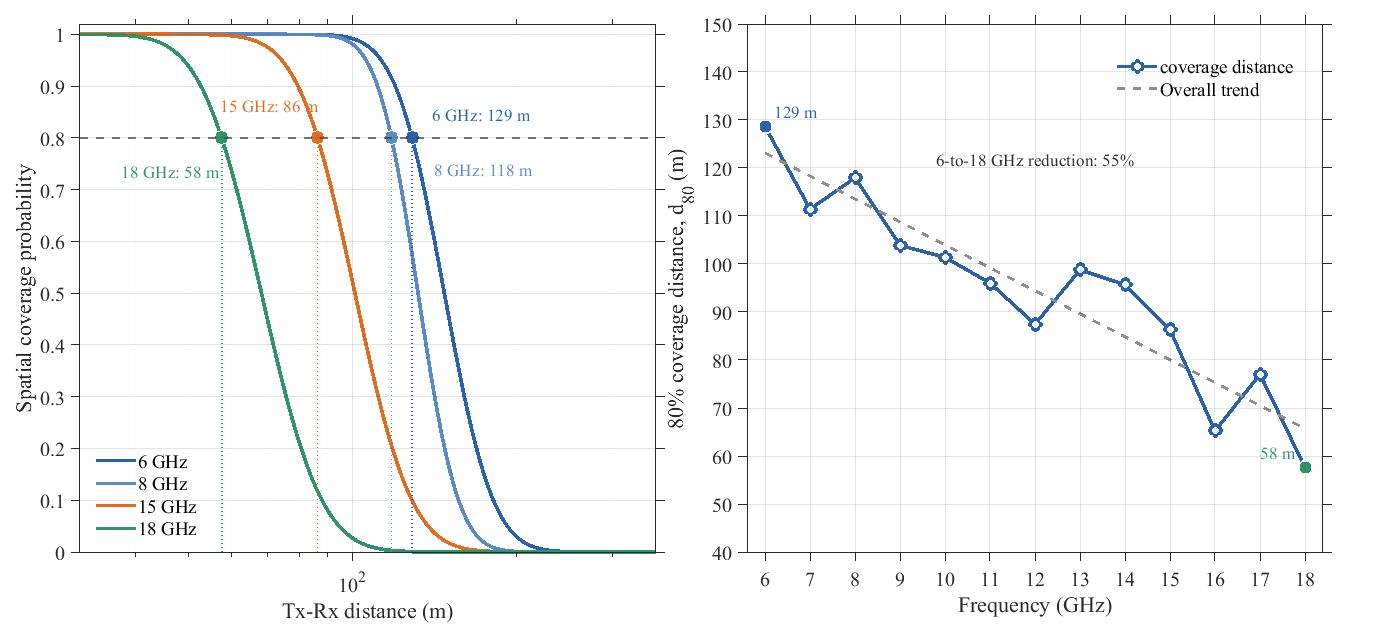}
\caption{Frequency evolution of measurement-based same-site coverage:
(a) spatial coverage probability at 6, 8, 15, and 18~GHz;
(b) 80\% coverage distance $d_{80}$ across 6--18~GHz.}
\label{fig:d80}
\end{figure}

The cross-frequency propagation penalty is further translated into same-site coverage performance using the criterion defined in Section~IV-A. As shown in Fig.~\ref{fig:d80}(a), the predicted $d_{80}$ values at 6, 8, 15, and 18~GHz are approximately 129, 118, 86, and 58~m, respectively, demonstrating an overall reduction in same-site coverage with increasing frequency. Fig.~\ref{fig:d80}(b) further shows the frequency evolution of $d_{80}$ across the 6--18~GHz band. Although $d_{80}$ generally decreases with frequency, local recoveries are observed at several frequencies, indicating that coverage is jointly affected by frequency-dependent FSPL, distance attenuation, and spatial variability. Overall, $d_{80}$ decreases from approximately 129~m at 6~GHz to 58~m at 18~GHz, corresponding to a reduction of approximately 55\%.

\begin{figure}[!hbt]
\centering
\includegraphics[width=\columnwidth]
{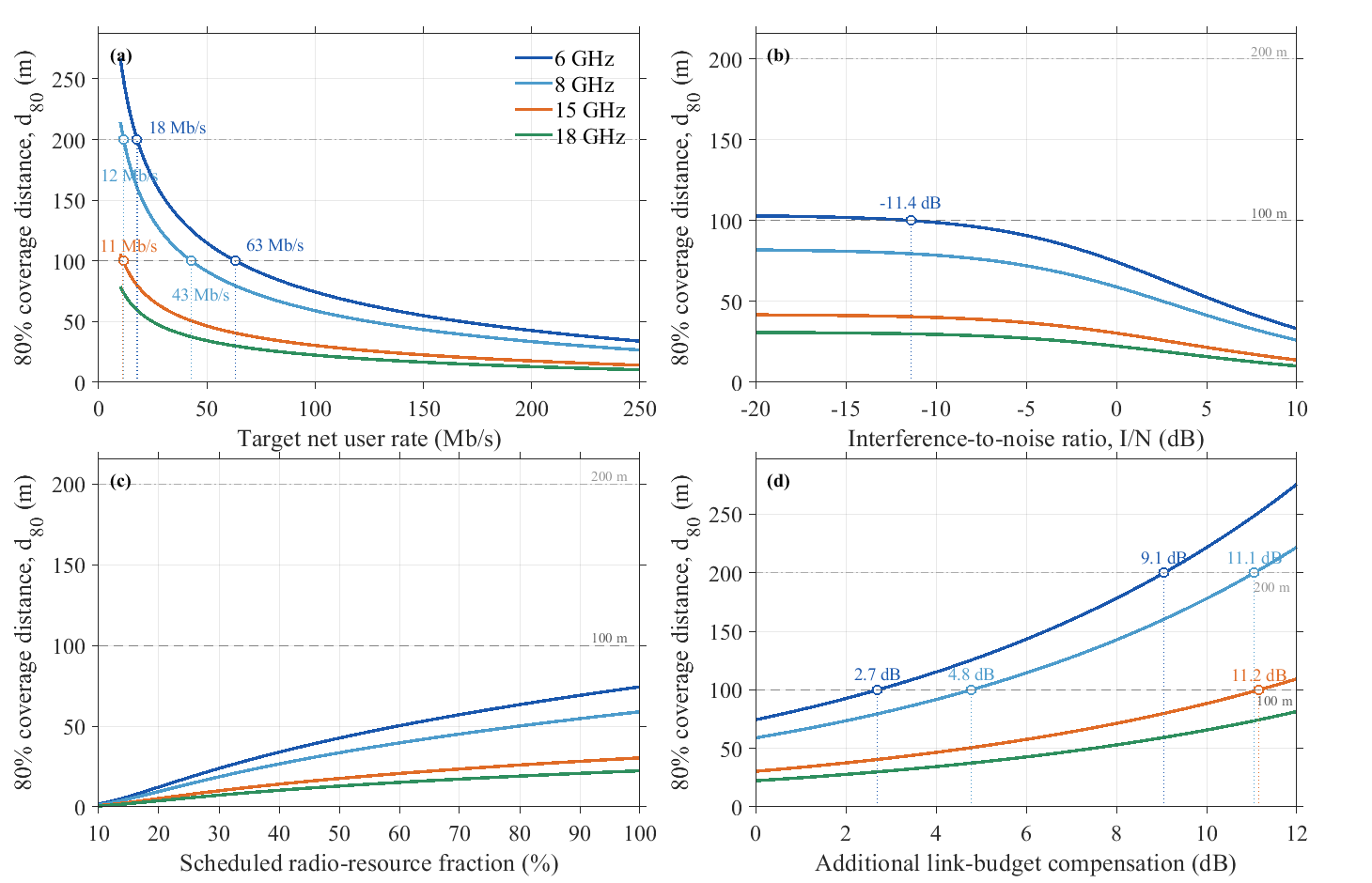}
\caption{Impact of service and system parameters on the 80\% same-site coverage distance $d_{80}$ at 6, 8, 15, and 18~GHz:
(a) target net user rate;
(b) interference-to-noise ratio;
(c) scheduled radio-resource fraction;
and (d) additional link-budget compensation.}
\label{fig:system_impact}
\end{figure}

Building on the baseline coverage results, the sensitivity of reliable same-site coverage to practical system constraints is further investigated. Fig.~\ref{fig:system_impact} evaluates the impact of target user rate, interference level, scheduled radio-resource fraction, and additional link-budget compensation on the 80\% reliable coverage distance $d_{80}$ at representative frequencies.

Fig.~\ref{fig:system_impact} is used to quantify how the measured frequency-dependent propagation differences are translated into practical coverage under different system constraints. Specifically, the target rate, interference level, and scheduled resource fraction modify the required SINR and hence the available path-loss budget, while additional link-budget compensation directly relaxes this constraint. The resulting $d_{80}$ therefore reflects the joint effect of the measured propagation parameters and the system operating condition. As shown in Fig.~\ref{fig:system_impact}(a)--(c), a higher target rate, stronger interference, or fewer scheduled resources reduces the available path-loss budget and consequently shortens $d_{80}$. More importantly, the separation between the frequency-dependent coverage curves persists under different system conditions, indicating that the observed coverage gap is jointly affected by frequency-dependent path loss, fitted distance attenuation, and spatial variability rather than in a particular system setting. Fig.~\ref{fig:system_impact}(d) further examines whether this propagation-induced disadvantage can be compensated by additional link budget. Although additional gain can substantially extend the coverage range, higher-frequency links still require more compensation to approach the low-frequency coverage level. These results show that the measured 6--18~GHz propagation differences impose a persistent coverage penalty, while system configuration determines how strongly this penalty appears in practical deployment.

\section{Conclusion}

This paper presented a wideband channel measurement campaign at 13 carrier frequencies measured at spatially aligned locations from 6 to 18~GHz in a UMa environment. The frequency evolution of PL, RMS DS, and PDP-based $K$-factor was characterized, showing increased propagation loss and a tendency toward more dominant-component-driven propagation at higher frequencies. Based on the measured PL, same-site coverage was further evaluated, showing that $d_{80}$ decreases from approximately 129~m at 6~GHz to 58~m at 18~GHz under the baseline configuration. The impacts of service rate, resource allocation, interference, and link-budget degradation were also quantified. These results provide measurement-based insights into the propagation and same-site deployment capability of 6--18~GHz wireless systems.

\section*{Acknowledgment}

This work was supported in part by Distinguished Young Scholars Continuation Program (62525101), National Natural Science Foundation of China (62571053, 62571059, and 62341128), Beijing Municipal Natural Fund (L243002), and BUPT-CMCC Joint Institute.

\end{document}